\documentclass{article}
\usepackage{hyperref}
\usepackage{algorithm}
\usepackage{algpseudocode}
\usepackage{graphicx}
\usepackage{amssymb}
\usepackage{url}

\newcommand{\ie}{i.e.\ }
\newcommand{\depend}[1]{\mathit{depend}(#1)}
\newcommand{\source}[1]{\mathit{source}(#1)}
\newcommand{\target}[1]{\mathit{target}(#1)}

\begin{document}

\title{A Mathematical Basis for\\
  the Chaining of Lossy Interface Adapters\footnote{This paper is a preprint of a paper accepted by IET Software and is subject to Institution of Engineering and Technology Copyright. When the final version is published, the copy of record will be available at IET Digital Library.}} 
\author{Yoo Chung \and Dongman Lee}
\maketitle

\begin{abstract}
  Despite providing similar functionality, multiple network services
  may require the use of different interfaces to access the
  functionality, and this problem will only get worse with the
  widespread deployment of ubiquitous computing environments.  One way
  around this problem is to use interface adapters that adapt one
  interface into another.  Chaining these adapters allows flexible
  interface adaptation with fewer adapters, but the loss incurred due
  to imperfect interface adaptation must be considered.  This paper
  outlines a matrix-based mathematical basis for analyzing the
  chaining of lossy interface adapters.  We also show that the problem
  of finding an optimal interface adapter chain is NP-complete with a
  reduction from 3SAT.
\end{abstract}

\section{Introduction}
\label{sec:introduction}

Similar network services can have different interfaces for providing
equivalent functionality, akin to the way there are a myriad of
infrared remote control protocols for televisions from different
manufacturers.  Multiple web services running over SOAP~\cite{soap}
may provide different interfaces for the same functionality, and the
same thing can happen for different embedded devices that essentially
do the same thing.  Even the same service from the same provider may
end up with different interfaces as newer versions are
developed~\cite{kaminski:cascon2006}.

One way to solve the problem of having a myriad of interfaces for the
same functionality is to standardize on a single interface.  This is
not always feasible due to economical or political considerations, so
another way is to develop and use adapters that can convert one
interface to another~\cite{gamma}.  This approach allows multiple
competing interfaces to coexist without constraining a network client
to one manufacturer or API standard, which would be required in
ubiquitous computing environments so as to allow a large number of
diverse computing devices interoperate with each other seamlessly.

The simplest way to adapt an unusable interface into a usable one is
to use a singe adapter to convert from one to the other.  This can
easily be extended so that multiple adapters are chained to adapt
interfaces~\cite{ponnekanti}, which can reduce the number of interface
adapters that are required.  However, it is not always feasible for an
adapter to perfectly convert one interface to another, since
interfaces are almost never created with conversion to other
interfaces in mind, and the problem is only worse when adapters are
chained~\cite{kim:percom2008}.

This paper outlines a mathematical basis for analyzing lossy interface
adaptation through the chaining of interface adapters.  In
section~\ref{sec:interface-adaptation}, we describe the background
behind interface adaptation.  Section~\ref{sec:math-basis} describes a
matrix-based mathematical basis for analyzing lossy interface adapter
chaining.  We show that optimal interface adapter chaining is
NP-complete in section~\ref{sec:complexity} using our mathematical
basis, so an exponential-time algorithm for the problem is suggested
in section~\ref{sec:greedy-algorithm}.  We discuss related work in
section~\ref{sec:related-work}, and the paper concludes in
section~\ref{sec:conclusions}.

\section{Interface adaptation}
\label{sec:interface-adaptation}

In this paper, we take the approach that services that provide similar
functionality can be accessed through different interfaces than that
used by the service itself through the use of pre-existing interface
adapters.  Each interface is accessed through methods, and an
interface adapter can provide an alternative interface by implementing
external methods through the methods available in the original
interface.  This is the same view as taken in
\cite{kim:percom2008}.\footnote{Contrary to how we call the original
  and converted interfaces as source and target interfaces,
  respectively, \cite{kim:percom2008} calls the original and converted
  interfaces as target and source interfaces, respectively.}

There can be various approaches to creating the interface adapters
themselves, from manual development of an adapter to automatic
generation through semantic or code analysis.  While manual
development of interface adapters is probably the most reliable
method, the mathematical framework described in this paper does not
preclude the use of alternative
methods~\cite{benatallah:caise2005,nezhad,zaha}, and the generation of
interface adapters themselves is outside the scope of this paper, as
our mathematical framework assumes a fixed set of interfaces and
pre-existing interface adapters.

\begin{figure}
  \centering
  \includegraphics[width=8cm]{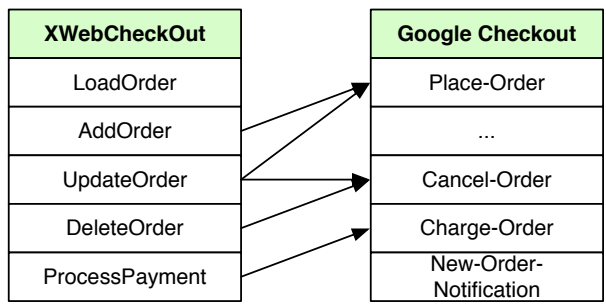}
  \caption{Example of service interface adaptation.}
  \label{fig:adapation-example}
\end{figure}

As a concrete example, we will describe how the web service
XWebCheckOut could be accessed using the Google Checkout API, an
example we based on one from \cite{nezhad}.  In
figure~\ref{fig:adapation-example}, we can see how XWebCheckOut has a
different interface from that of Google Checkout. For a network client
that only knows how to use the Google Checkout API, it would need an
adapter which can convert the source interface for XWebCheckOut to the
target interface for Google Checkout.

A developer could implement methods for the Google Checkout interface
by using methods available in the XWebCheckOut interface.  For
example, the \textsc{Place-Order} method in the Google Checkout
interface could be implemented using the \textsc{AddOrder} and
\textsc{UpdateOrder} methods in the XWebCheckOut interface.  Doing
this for each method in the Google Checkout interface will result in
an interface adapter that adapts the XWebCheckout interface to the
Google Checkout interface.

However, interface adaptation might not be perfect as some methods in
the target interface simply cannot be implemented using only methods
in the source interface, resulting in lossy interface adaptation.  We
can see this in figure~\ref{fig:adapation-example}, where there is no
feasible way to implement the \textsc{New-Order-Notification} method
using only methods available from the XWebCheckOut interface, assuming
that the method cannot be implemented independently of XWebCheckOut.

Since a network client is using the target interface to access a
service, the lossiness in the target interface is of more interest
than the inability to provide access to the full range of
functionality provided in the source interface.  For example, for a
network client that only knows how to use the Google Checkout API, it
is more relevant that an interface adapter may not be able to provide
the \textsc{New-Order-Notification} method in the Google Checkout
interface, rather than that the functionality provided by the
\textsc{LoadOrder} method in the XWebCheckOut interface is missing.

If we require that all available interfaces for similar services must
be adapted between each other with only a single adapter in between,
then the number of interface adapters required is in the order of
$n^2$.  Developing all the required adapters can be impractical, so
interface adapters can be \emph{chained} to adapt a source interface
to one interface, this interface adapted to another interface, and so
on until we get a target interface that a network client knows how to
use~\cite{ponnekanti}.  In the best case, we can even get away with
only $n$~adapters given $n$~interfaces.

\begin{figure}
  \centering
  \includegraphics[width=10cm]{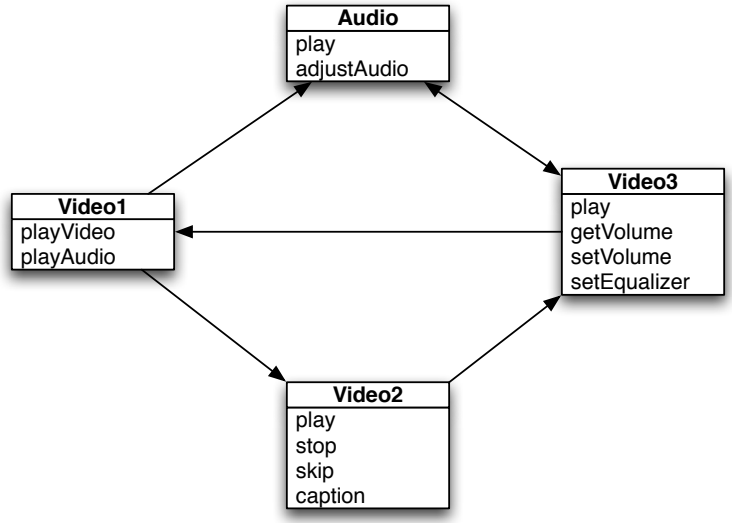}
  \caption{Multiple interfaces related by interface adapters.}
  \label{fig:multiple-interface-example}
\end{figure}

However, different chains of interface adapters result in different
lossiness in the interface adaptation, so we need a way to analyze the
chaining of lossy interface adapters.  We will look at another example
in figure~\ref{fig:multiple-interface-example}, where there are four
interfaces and six interface adapters, each of the latter represented
by an arrow from the source interface to the target interface it
converts from and to.

Each interface may have the following characteristics:
\begin{itemize}
\item \textit{Video1} can play both video and audio files.
\item \textit{Video2} can only play video files, but can stop
  playback, skip over a fixed amount of time, and select captions.
\item \textit{Video3} can only play video files, but can get and set
  the volume and set the equalizer for the audio output.
\item \textit{Audio} can only play video files, but can set audio
  properties, which are the volume and the equalizer.
\end{itemize}

And each interface adapter may be implemented as in the following,
where methods in a target interface not mentioned are not adapted to:
\begin{itemize}
\item The adapter \textit{Video1toVideo2}, which converts
  \textit{Video1} to \textit{Video2}, may implement the \textit{play}
  method of \textit{Video2} by using the \textit{playVideo} method of
  \textit{Video1}.
\item The adapter \textit{Video2toVideo3}, which converts
  \textit{Video2} to \textit{Video3}, may implement the \textit{play}
  method of \textit{Video3} by using the \textit{play} method of
  \textit{Video2}.
\item The adapter \textit{Video1toAudio}, which converts
  \textit{Video1} to \textit{Audio}, may implement the \textit{play}
  method of \textit{Audio} by using the \textit{playAudio} method of
  \textit{Video1}.
\item The adapter \textit{AudiotoVideo3}, which converts
  \textit{Audio} to \textit{Video3}, may implement the
  \textit{setVolume} and \textit{setEqualizer} methods of
  \textit{Video3} by using the \textit{adjustAudio} method of
  \textit{Audio}.\footnote{While it may seem odd to have an adapter
    from \textit{Audio} to \textit{Video3} which cannot adapt a
    playback method, it can still be useful when someone wishes to
    reduce loud noises from an audio device with interface
    \textit{Audio} using a remote control that only understands the
    interface \textit{Video3}.}
\item The adapter \textit{Video3toAudio}, which converts
  \textit{Video3} to \textit{Audio}, may implement the
  \textit{adjustAudio} method of \textit{Audio} by using the
  \textit{setVolume} and \textit{setEqualizer} methods of
  \textit{Video3}.
\item The adapter \textit{Video3toVideo1}, which converts
  \textit{Video3} to \textit{Video1}, may implement the
  \textit{playVideo} method of \textit{Video1} using the \textit{play}
  method of \textit{Video3}.
\end{itemize}

A service with interface \textit{Video1} may be available, and we may
want to access it using a client that only understands interface
\textit{Video3}.  There is no interface adapter which directly
converts interface \textit{Video1} to \textit{Video3}, but there are
interface adapter chains which can indirectly do the conversion.
Chaining \textit{Video1toVideo2} with \textit{Video2toVideo3} or
chaining \textit{Video1toAudio} with \textit{AudiotoVideo3} can
convert interface \textit{Video1} to \textit{Video3}.

Given multiple possible interface adapter chains, we would want to use
the best interface adapter chain that can provide the most methods in
the target interface.  Associating a cost with an adapter depending on
how well it adapts the methods in its target interface and using
minimum-cost path algorithms such as Dijkstra's
algorithm~\cite{dijkstra:mathematik1959} would be an obvious approach
to choose the best interface adapter chain.  However, this naive
approach would not work as we will see from the example in
figure~\ref{fig:multiple-interface-example}.

\textit{Video1toAudio} and \textit{AudiotoVideo3} can adapt one out of
two methods in \textit{Audio} and \textit{Video3}, respectively.  In
contrast, \textit{Video1toVideo2} and \textit{Video2toVideo3} can
adapt one out of four methods in \textit{Video2} and \textit{Video3}.
One might think that the \textit{Video1toAudio} and
\textit{AudiotoVideo3} chain would be better than the
\textit{Video1toVideo2} and \textit{Video2toVideo3} chain simply by
looking at how lossy each interface adapter is, but one would be
wrong.

\textit{AudiotoVideo3} requires the \textit{adjustAudio} method in
\textit{Audio} to implement the \textit{setVolume} and
\textit{setEqualizer} methods in \textit{Video3}.  However,
\textit{Video1toAudio} cannot implement the \textit{adjustAudio}
method in \textit{Audio}, so the \textit{Video1toAudio} and
\textit{AudiotoVideo3} chain ends up with no available methods for
\textit{Video3}.  In contrast, the \textit{Video1toVideo2} and
\textit{Video2toVideo3} chain can provide the method \textit{play} for
\textit{Video3}.  A single number for each interface adapter cannot
express such dependencies properly, so we need a more precise approach
to analyze the lossiness in interface adapter chains.

In the rest of the paper, we will discuss how to mathematically
analyze the lossiness incurred from the chaining of interface
adapters.  We will also assume that interface adapters are implemented
as transparently as possible: while an interface adapter may not be
able to provide all of the methods in the target interface, the
methods it will provide will work just as if they were invoked
directly on a service having the target interface.

\section{Mathematical basics}
\label{sec:math-basis}

We can start formalizing the problem of lossy interface adaptation by
defining an \emph{interface adapter graph}.  This is a directed graph
where interfaces are nodes and adapters are edges.  If there are
interfaces $I_1$ and $I_2$ with an adapter $A$ that adapts source
interface $I_1$ to target interface $I_2$, then $I_1$ and $I_2$ would
be nodes in the interface adapter graph while $A$ would be a directed
edge from $I_1$ to $I_2$.

We do not assume that there can be at most one adapter which can adapt
one interface to another.  This reflects the fact that there can be
multiple adapters from different developers, similar to how there can
be multiple device drivers available for a graphics card.  It also
simplifies some of the arguments, although they would still hold even
with such a restriction with only minor changes in the proofs.

We will be using a range convention for the index notation used to
express matrixes and vectors~\cite{index-notation}.

\subsection{Method dependencies}
\label{sec:dependency-matrix}

The next step is to formally describe each adapter, \ie each edge in
the interface adapter graph, in a way that would be useful for
analyzing lossiness.  We should be able to figure out which methods in
the target interface can be provided by an interface adapter given the
methods available in the source interface.  We do this by defining a
\emph{method dependency matrix}, a boolean matrix which describes how
an interface adapter implements methods in the target interface using
available methods in the source interface.

The method dependency matrix $a_{ji}$ for an adapter~$A$, where
$a_{ji}$ represents either the matrix itself or a single component in
the matrix depending on the context, is defined by how the adapter
depends on the availability of a method in the source interface in
order to implement a method in the target interface.  $a_{ji}$ is true
if and only if method~$j$ in the target interface can be implemented
only if method~$i$ in the source interface is available.  We denote
the method dependency matrix associated with an adapter~$A$ as
$\depend{A}$.

We also define a \emph{method availability vector} $p_i$ for an
interface, where each component $p_i$ is true if and only if
method~$i$ is available.  This boolean vector is not intrinsic to an
interface, unlike the method dependency matrix which \emph{is}
intrinsic to an interface adapter.  Instead, it is used to represent
the lossiness in interface adaptation such that method~$i$ in the
target interface can be used only if $p_i$ is true.  For a fully
functional service that implements all methods specified in its
interface, the components of its method availability vector should all
be true.  We denote the number of true components in method
availability vector~$p_i$ as $\|p_i\|$, which is equivalent to the
Manhatten norm~\cite{manhattan-norm} when true and false components
are replaced by 1 and 0, respectively.

Given method availability vector $p_i$ for a source interface and the
method dependency matrix $a_{ji}$ for an interface adapter, we can
derive the method availability vector $q_j$ for the target interface.
A method~$j$ in the target interface can only be implemented if all of
the methods it depends on are available in the source interface.  So
if $q_j$ is to be true for fixed $j$, then all $p_i$ must be true when
$a_{ji}$ is true:
\begin{equation}
  \label{eq:dependency}
  q_j = \bigwedge_i ( a_{ji} \rightarrow p_i)
  = \bigwedge_i ( \neg a_{ji} \vee p_i)
\end{equation}

However, equation~(\ref{eq:dependency}) is incomplete in that it does
not properly distinguish between methods which can always be
implemented and methods which cannot be implemented given the source
interface.  For example, a method that returns the value of $\pi$ does
not need anything from the source interface, whereas there would be no
way to implement a video playback method given only a source interface
specialized exclusively for audio playback.  For both cases, all
$a_{ji}$ are false for a specific method~$j$, and
equation~(\ref{eq:dependency}) would give the wrong result for the
latter case.

This can be worked around by defining a \emph{dummy method} that is
never available for every interface.  We arbitrarily call this
``method~1'', so that $p_1$ will always be false for any method
availability vector.  It is easy to see that extending the definition
of the method dependency matrix with the following rules is consistent
with our definitions and equations for the method dependency matrix
and method availability vector:
\begin{itemize}
\item $a_{11}$ is true, while $a_{1i}$ is set to false for all $i \neq
  1$.
\item If method~$j$ can always be implemented in the target interface,
  set $a_{ji}$ to false for all $i$.
\item If method~$j$ can never be implemented given the source
  interface, set $a_{j1}$ to true, while $a_{ji}$ is set to false for
  all $i \neq 1$.
\item If method~$j$ depends on the availability of actual methods in
  the source interface, then $a_{j1}$ is false.
\end{itemize}

For succintness, we denote a method availability vector for
interface~$I$ which represents that all methods are available, \ie
when the component for the dummy method is false while all the other
components are true, by $\mathbf{1}'_I$.

We also define the operator $\otimes$ for a method dependency matrix
as applied to a method availability vector to represent the operation
in equation~(\ref{eq:dependency}), or in other words:
\begin{equation}
  \label{eq:operator}
  a_{ji} \otimes p_i \equiv \bigwedge_i (\neg a_{ji} \vee p_i)
\end{equation}

It is easy to see that a square boolean matrix where the diagonals are
true and the rest of the components are false is an identity matrix
for the adaptation operator $\otimes$.  We denote an identity matrix
with $n$~rows as $\mathbf{I}_n$.

\subsection{Adapter composition}
\label{sec:adapter-composition}

To analyze the chaining of lossy interface adapters, we are also
interested in how to derive a composite method dependency matrix from
the composition of two method dependency matrixes, which would be
equivalent to describing the chaining of two interface adapters as if
they were a single interface adapter.

Given interfaces $I_1$, $I_2$, and $I_3$, let the corresponding method
availability vectors be $p_i$, $q_j$, and $r_k$.  In addition, let
there be interface adapters $A_1$ and $A_2$, where $A_1$ converts
$I_1$ to $I_2$ and $A_2$ converts $I_2$ to $I_3$, with corresponding
method dependency matrixes $a_{ji}$ and $b_{kj}$, respectively.  We
would like to know how to derive the method dependency matrix that
would correspond to an interface adapter equivalent to $A_1$ and $A_2$
chained together.

From equation~(\ref{eq:dependency}) and our assumptions:
\begin{eqnarray*}
  r_k &=& \bigwedge_j ( \neg b_{kj} \vee q_j ) \\
  &=& \bigwedge_j \left( \neg b_{kj} \vee \bigwedge_i ( \neg
      a_{ji} \vee p_i ) \right) \\ 
  &=& \bigwedge_j \bigwedge_i ( \neg b_{kj} \vee \neg a_{ji} \vee p_i ) \\
  &=& \bigwedge_i \bigwedge_j ( \neg b_{kj} \vee \neg a_{ji} \vee p_i ) \\
  &=& \bigwedge_i \left( \bigwedge_j ( \neg b_{kj} \vee \neg a_{ji} ) 
    \vee p_i \right) \\
  &=& \bigwedge_i \left( \neg \bigvee_j ( b_{kj} \wedge a_{ji} ) \vee
    p_i \right)
\end{eqnarray*}

We reuse the operator $\otimes$ to represent the composition of two
method dependency matrixes, and by comparing the above with
equation~(\ref{eq:dependency}), we can define it as:
\begin{equation}
  \label{eq:composition}
  b_{kj} \otimes a_{ji} = \bigvee_j (b_{kj} \wedge a_{ji})
\end{equation}

$\mathbf{I}_n$ from section~\ref{sec:dependency-matrix} is also an
identity matrix for the method dependency matrix composition operator
$\otimes$.

The $\otimes$ operator is ``associative''\footnote{It is not
  technically associative in this context as the $\otimes$ operator as
  applied to method dependency matrixes is not really the same as the
  $\otimes$ operator as applied to a method dependency matrix and a
  method availability vector, similarly to how $\times$ for numbers is
  different from $\times$ for sets.} when applied to method dependency
matrixes and a method availability vector, \ie \(b_{kj} \otimes
(a_{ji} \otimes p_i) = (b_{kj} \otimes a_{ji}) \otimes p_i\), which
shows that in terms of lossiness, chaining adapters and then applying
it to a source interface is equivalent to applying each adapter one by
one to the source interface:
\begin{eqnarray*}
  b_{kj} \otimes (a_{ji} \otimes p_i) 
  &=& \bigwedge_j \left( \neg b_{kj} \vee \bigwedge_i ( \neg a_{ji}
    \vee p_i ) \right) \\
  &=& \bigwedge_j \bigwedge_i ( \neg b_{kj} \vee \neg a_{ji} \vee p_i ) \\
  &=& \bigwedge_i \bigwedge_j ( \neg b_{kj} \vee \neg a_{ji} \vee p_i ) \\
  &=& \bigwedge_i \bigwedge_j ( \neg (b_{kj} \wedge a_{ji}) \vee p_i ) \\
  &=& \bigwedge_i \left( \bigwedge_j \neg (b_{kj} \wedge a_{ji})
    \vee p_i \right) \\
  &=& \bigwedge_i \left( \neg \bigvee_j (b_{kj} \wedge a_{ji}) 
    \vee p_i \right) \\
  &=& (b_{kj} \otimes a_{ji}) \otimes p_i
\end{eqnarray*}

Likewise, method dependency matrix composition is associative:
\begin{eqnarray*}
  c_{lk} \otimes (b_{kj} \otimes a_{ji})
  &=& \bigvee_k \left( c_{lk} \wedge \bigvee_j (b_{kj} \wedge a_{ji}) \right) \\
  &=& \bigvee_k \bigvee_j (c_{lk} \wedge b_{kj} \wedge a_{ji}) \\
  &=& \bigvee_j \bigvee_k (c_{lk} \wedge b_{kj} \wedge a_{ji}) \\
  &=& \bigvee_j \left( \bigvee_k (c_{lk} \wedge b_{kj}) \wedge a_{ji}) \right) \\
  &=& (c_{lk} \otimes b_{kj}) \otimes a_{ji}
\end{eqnarray*}

However, method dependency matrix composition is \emph{not}
commutative, as can be easily seen by considering the composition of
method dependency matrixes that are not square matrixes.

We can also formalize the somewhat vague intuition that a longer
interface adapter chain is worse in terms of lossiness.  If $A_1$ and
$A_2$ are interface adapters, where $A_1$ converts $I_1$ to $I_2$ and
$A_2$ converts $I_2$ to $I_3$, with $a_{ji} = \depend{A_1}$ and
$b_{kj} = \depend{A_2}$ in which $a_{11}$ and $b_{11}$ are both true
as in section~\ref{sec:dependency-matrix}, then for $p_i = b_{kj}
\otimes \mathbf{1}'_{I_2}$ and $p'_i = b_{kj} \otimes a_{ji} \otimes
\mathbf{1}'_{I_1}$:
\begin{displaymath}
  p_k = (\neg b_{k1} \vee f) \wedge \bigwedge_{j \neq 1} (\neg
  b_{kj} \vee t) = \neg b_{k1}
\end{displaymath}
\begin{eqnarray*}
  p'_k &=& \bigwedge_j \left( \neg b_{kj} \vee \left( (\neg a_{j1}
      \vee f) \wedge \bigwedge_{i \neq 1} (\neg a_{ji} \vee t) \right)\right) \\
  &=& \bigwedge_j (\neg b_{kj} \vee \neg a_{j1}) \\
  &=& \neg b_{k1} \wedge \bigwedge_{j \neq 1} (\neg b_{kj} \vee a_{j1})
\end{eqnarray*}

\begin{equation}
  \label{eq:monotonicity}
  \therefore p'_k \rightarrow p_k
\end{equation}

With $I_1$ and $I_2$ being the source interfaces for the interface
adapters that $a_{ji}$ and $b_{kj}$ represent, respectively, we can
also infer from equation~(\ref{eq:monotonicity}) that
\begin{equation}
  \label{eq:numeric-monotonicity}
  \|b_{kj} \otimes \mathbf{1}'_{I_2}\|
  \geq
  \|b_{kj} \otimes a_{ji} \otimes \mathbf{1}'_{I_1}\|
\end{equation}
which, along with the associativity of method dependency matrix
composition, formalizes the notion that extending an interface adapter
chain is worse in terms of lossiness.

The definitions of the method dependency matrix and the method
availability vector in section~\ref{sec:dependency-matrix}, along with
the associativity rules proven in this section, provide a succinct way
to mathematically express and analyze the chaining of lossy interface
adapters.

\subsection{An example}
\label{sec:basis-example}

\begin{figure}
  \centering
  \includegraphics[width=7cm]{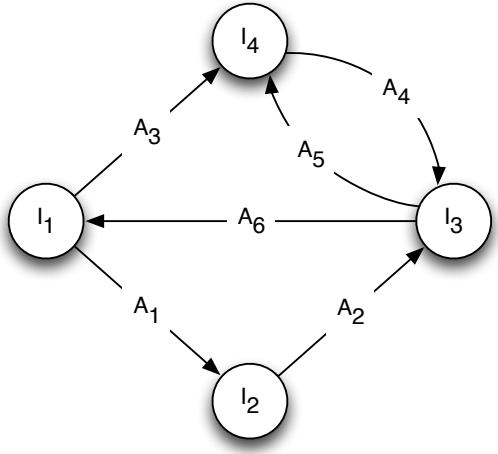}
  \caption{Interface adapter graph for figure~\ref{fig:multiple-interface-example}.}
  \label{fig:adapter-graph-example}
\end{figure}

As an example, we apply the mathematical framework to the interfaces
and adapters in figure~\ref{fig:multiple-interface-example}.  We will
denote interfaces \textit{Video1}, \textit{Video2}, \textit{Video3},
and \textit{Audio} as $I_1$, $I_2$, $I_3$, and $I_4$, respectively,
while $A_1$, $A_2$, $A_3$, $A_4$, $A_5$, and $A_6$ denote the
interface adapters \textit{Video1toVideo2}, \textit{Video2toVideo3},
\textit{Video1toAudio}, \textit{AudiotoVideo3},
\textit{Video3toAudio}, and \textit{Video3toVideo1}, respectively.  We
also index each method in the order they appear in
figure~\ref{fig:multiple-interface-example} along with an extra dummy
method with index 1, and let $a^k_{ji} = \depend{A_k}$.
Figure~\ref{fig:multiple-interface-example} is already an interface
adapter graph, which is simplified and labeled in
figure~\ref{fig:adapter-graph-example}.

Some method dependency matrixes would be:
\begin{displaymath}
  a^1_{ji} =
  \left(
    \begin{array}{ccc}
      t & f & f \\
      f & t & f \\
      t & f & f \\
      t & f & f \\
      t & f & f
    \end{array}
  \right)
\end{displaymath}
\begin{displaymath}
  a^2_{ji} =
  \left(
    \begin{array}{ccccc}
      t & f & f & f & f \\
      f & t & f & f & f \\
      t & f & f & f & f \\
      t & f & f & f & f \\
      t & f & f & f & f
    \end{array}
  \right)
\end{displaymath}
\begin{displaymath}
  a^5_{ji} =
  \left(
    \begin{array}{ccccc}
      t & f & f & f & f \\
      t & f & f & f & f \\
      f & f & f & t & t
    \end{array}
  \right)
\end{displaymath}

Given a fully functional service which conforms to interface
\textit{Video1}, we would expect that only the \textit{play} method
would be available for interface \textit{Video3} after going through
the adapter chain $A_1$ and $A_2$, which can be verified by computing
the method availability vector $a^2_{kj} \otimes a^1_{ji} \otimes
\mathbf{1}'_{I_1}$:
\begin{displaymath}
  a^2_{kj} \otimes a^1_{ji} \otimes \mathbf{1}'_{I_1} = [f, t, f, f, f]
\end{displaymath}

One can also verify the following by hand, which would be expected
from the associativity of $\otimes$.  Associativity can be very useful
in developing algorithms analyzing chains of lossy interface adapters,
since fragments of an interface adapter chain can be assembled
independently and still give the same method dependency matrix for the
whole chain.
\begin{eqnarray*}
  \lefteqn{a^5_{lk} \otimes a^2_{kj} \otimes a^1_{ji} \otimes \mathbf{1}'_{I_1}} \\
  &=&  a^5_{lk} \otimes (a^2_{kj} \otimes (a^1_{ji} \otimes \mathbf{1}'_{I_1})) \\
  &=&  ((a^5_{lk} \otimes a^2_{kj}) \otimes a^1_{ji}) \otimes \mathbf{1}'_{I_1} \\
  &=&  (a^5_{lk} \otimes a^2_{kj}) \otimes (a^1_{ji} \otimes \mathbf{1}'_{I_1}) \\
  &=& [f, f, f]
\end{eqnarray*}

We can also verify the following, which is consistent with equations
(\ref{eq:monotonicity}) and (\ref{eq:numeric-monotonicity}), and is in
line with the intuition that extending an adapter chain can only be
worse in terms of lossiness, although this does not mean that a longer
adapter chain is always worse than a shorter adapter chain.
\begin{eqnarray*}
  a^5_{lk} \otimes \mathbf{1}'_{I_3} &=& [f, f, t] \\
  a^5_{lk} \otimes a^2_{kj} \otimes \mathbf{1}'_{I_2} &=& [f, f, f]
\end{eqnarray*}

\section{Optimal adapter chaining}
\label{sec:complexity}

One of the things that could be hoped from the mathematical framework
in section~\ref{sec:math-basis} is that it could help with the
development of an efficient algorithm for optaining an optimal
interface adapter chain from an actual service to a target interface
that incurs the least loss in terms of functionality.  Unfortunately,
the problem is NP-complete, as will be shown in this section, dashing
hopes for such an algorithm.

First, we must formally describe the problem, which we will call
CHAIN.  Let us have an interface adapter graph $(\{I_i\}, \{A_i\})$,
where $\{I_i\}$ is the set of interfaces and $\{A_i\}$ is the set of
interface adapters.  Let $a^k$ be the method dependency matrix
associated with adapter~$A_k$.  Let $S \in \{I_i\}$ be the source
interface and $T \in \{I_i\}$ be the target interface.  Then the
problem is whether there is an interface adapter chain $[A_{P(1)},
A_{P(2)}, \ldots, A_{P(m)}]$ such that the source of $A_{P(1)}$ is
$S$, the target of $A_{P(m)}$ is $T$, and $\|v^T\| = \|a^{P(m)}
\otimes \cdots \otimes a^{P(2)} \otimes a^{P(1)} \otimes
\mathbf{1}'_S\|$ is at least as large as some parameter $N$.

Informally, this is an optimization problem which tries to maximize
the number of methods that can be used in a fixed target interface,
obtained by applying an interface adapter chain on a fully-functional
service which conforms to the source interface.  We show that the
problem is NP-complete by reducing 3SAT~\cite{cook} to CHAIN.

Based on the conjunctive normal form of a boolean expression $E$ with
exactly 3 literals in each clause, we will construct an interface
adapter graph $G$ in three parts and the corresponding method
dependency matrixes.  One part will model the setting of each variable
to true or false, another part will model the value of each clause
once the variable values are set, and the last part will serve as a
filter so that $E$ is satisfiable if and only if there is a chain in
$G$ such that $\|v^T\|$ equals the number of clauses in $E$.

\begin{figure}
  \centering
  \includegraphics[width=10cm]{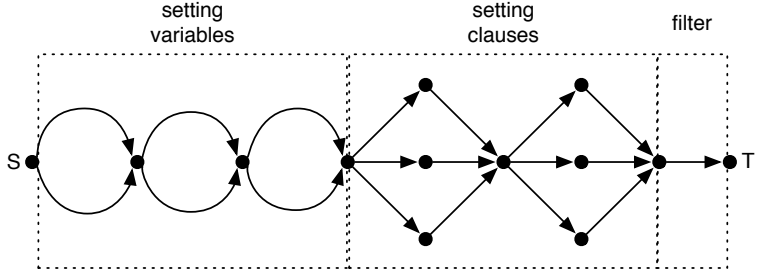}
  \caption{General form of a boolean expression reduced to an
    interface adapter graph.}
  \label{fig:reduction-generic-example}
\end{figure}

Figure~\ref{fig:reduction-generic-example} shows what a reduction from
an instance of 3SAT to an instance of CHAIN would generally look like.

\subsection{Representing values}
\label{sec:representation}

We will represent values of literals and clauses using the method
availability vector for each interface, where all but one of the nodes
in the constructed interface adapter graph will contain the same set
of methods.  At certain points in the interface adapter graph, a true
or false component in the method availability vector would directly
map to the value of a literal or a clause.

For almost all nodes, including the source, the set of methods will be
fixed with one dummy method, one method for each clause, and one
method for each literal, so almost all method dependency matrixes will
be square matrixes.  As the method dependency matrixes will have large
parts in common with the identity matrix, we will only be mentioning
how each matrix differs from the identity matrix.

Each method will be labeled as follows:
\begin{itemize}
\item The dummy method will be labeled $d$.
\item For each clause $c_i$, the method will be labeled $c_i$.
\item For each variable $v_i$, the method for the variable itself will
  be labeled $l_i$, while the method for the negation of the variable
  will be labeled $\overline{l_i}$.
\end{itemize}

There is a single method dependency matrix used in the filter part of
the graph that will not be a square matrix.

\subsection{Handling literals}
\label{sec:handling-literals}

The basic approach of this part of the graph, which we will call the
\emph{variable handling subgraph}, is to set the value for each
variable depending on which adapters are chosen to be included in the
chain.  For each variable $v_1$, $v_2$, \ldots, $v_v$, we define nodes
$V_1$, $V_2$, \ldots, $V_v$, and we let $V_0 = S$.  Between each
$V_{i-1}$ and $V_i$, we define two adapters which will leave
everything about the method availability vector unchanged from one
node to the next except for the components corresponding to the
literals for $v_i$.  One will make the variable effectively true,
while the other will make the variable effectively false.

For each $V_i$ for $i > 0$, we will define a \emph{positive literal
  adapter} $A_{l_i}$ with method dependency matrix $a^{l_i}$ and a
\emph{negative literal adapter} $A_{\overline{l_i}}$ with method
dependency matrix $a^{\overline{l_i}}$.  For the positive literal
adapter, $a^{l_i}_{l_i j}$ is false for all $j$,
$a^{l_i}_{\overline{l_i} d}$ is true, and $a^{l_i}_{\overline{l_i} j}$
is false for all $j$ other than $d$.  Similarly for the negative
literal adapter, $a^{\overline{l_i}}_{\overline{l_i} j}$ is false for
all $j$, $a^{\overline{l_i}}_{l_i d}$ is true, and
$a^{\overline{l_i}}_{l_i j}$ is false for all $j$ other than $d$.

It should then be easy to see that for a method availability vector
$p_i$ with a false $p_d$, all components of $a^{l_i} \otimes p_i$
should be the same as $p_i$ except for the components corresponding to
$l_i$ and $\overline{l_i}$, which will be true and false,
respectively.  Likewise, all components of $a^{\overline{l_i}} \otimes
p_i$ should be the same as $p_i$ except for the components
corresponding to $l_i$ and $\overline{l_i}$, which will be false and
true, respectively.

\begin{figure}
  \centering
  \includegraphics[width=5cm]{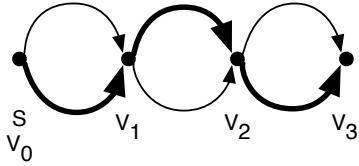}
  \caption{Choosing a variable assignment.}
  \label{fig:variable-single-path}
\end{figure}

The rest of the interface adapter graph will be the descendant of
$V_v$, so any adapter chain from $S$ to $T$ \emph{must} go through all
of $V_0$, $V_1$, \ldots, $V_v$ in order, and for every variable one
and only one of the positive literal adapter or the negative literal
adapter must be chosen as in figure~\ref{fig:variable-single-path} due
to the structure of the variable handling subgraph.  This is
equivalent to choosing a variable assignment, and at $V_v$, the method
availability vector $p_i$ will be such that for each variable~$v_i$,
$p_{l_i}$ and $p_{\overline{l_i}}$ will have opposite values, so that
it would be the same as setting the value of $v_i$ to $p_{l_i}$.

\subsection{Handling clauses}
\label{sec:handling-clauses}

Based on the variable assignment that is taken care of by the variable
handling subgraph in section~\ref{sec:handling-literals}, this part of
the graph, which we will call the \emph{clause handling subgraph}, is
responsible for determining the value of each clause.

In order to model disjunction, not only do we define a node $C_i$ for
each clause~$c_i$, we also define three subnodes $C_{ij}$, for $j$
from 1 to 3, for each of the literals in the clause.  These nodes are
separate from those defined in section~\ref{sec:handling-literals}.
The idea is that if any of the literals are true, then at least one of
the nodes will end up with a method availability vector marking the
clause as true, so we can use this to mark the same for $C_i$ itself.
We also set $C_0 = V_v$ for convenience of notation, and $c$ will be
the number of clauses.

For each clause $c_i$, there are edges from $C_{i-1}$ to each of the
subnodes $C_{ij}$, and in turn there are edges from each subnode
$C_{ij}$ to $C_i$.  So there will be three alternate paths from
$C_{i-1}$ to $C_i$.

For edge $(C_{i-1}, C_{ij})$, if $l$ corresponds to the literal for
$C_{ij}$, the method dependency matrix $a$ for the edge is defined by
setting $a_{c_i l}$ to true and $a_{c_i k}$ to false for all $k$ other
than $l$.  Then it should be easy to see that $a \otimes p$ is the
same as the method availability vector $p$ except for the component
$p_{c_i}$, which would be true if and only if $p_l$ is also true.  For
edge $(C_{ij}, C_i)$, the corresponding method dependency matrix is
simply the identity matrix.

If clause~$c_i$ is true, then one of the literals must be true.  Then
the path through the subnode $C_{ij}$ for the true literal will result
in a true component for the clause in the method availability vector
at $C_i$.  If the clause is not true, then the same component will be
false no matter the path taken, since it will be false for all
subnodes $C_{ij}$.

\begin{figure}
  \centering
  \includegraphics[width=5cm]{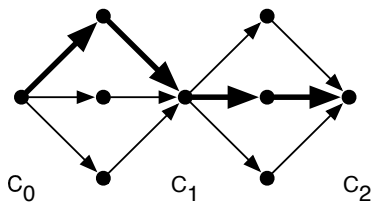}
  \caption{Disjunction through alternate paths.}
  \label{fig:clause-single-path}
\end{figure}

$T$ will be the descendant of $C_c$, and since the source is in the
variable handling subgraph, which is only connected to the clause
handling subgraph by $C_0$, any interface adapter chain from $S$ to
$T$ \emph{must} go through each of the nodes $C_0$, $C_1$, \ldots,
$C_c$ in order as in figure~\ref{fig:clause-single-path}.  And if all
clauses are true with the variable assignment done in the variable
handling subgraph, which is equivalent to choosing which adapters to
include from the subgraph, only then will there be a path from $C_0$
to $C_c$ which will result in true components for all clauses in the
method availability vector at $C_c$.

\subsection{Filtering}
\label{sec:filtering}

The last part of the constructed interface adapter graph is the
filtering part, which discards all methods corresponding to literals
from the method availability vector so that only the dummy method and
methods corresponding to clauses remain.

The filtering subgraph is made up of only two nodes and a single edge.
One of the nodes is the target~$T$, and its interface only contains
the dummy method and all the methods corresponding to clauses.  The
other node is $C_c$ from section~\ref{sec:handling-clauses}.  The
$(c+1) \times (2v+c+1)$ method dependency matrix $a_{ji}$ for the edge
from $C_c$ to $T$ defined as follows accomplishes the filtering:
\begin{itemize}
\item For all clauses $c_i$, $a_{c_i c_i}$ is true.
\item For the dummy method, $a_{dd}$ is true.
\item All other compenents are false.
\end{itemize}

\subsection{Analysis of the reduction}
\label{sec:reduction-analysis}

The constructed interface adapter graph has $v+4c+2$~nodes and
$2v+6c+1$~edges, where $v$ is the number of variables and $c$ is the
number of clauses.  Also, each method dependency matrix has at most
$(1+c+2v)^2$ components, so the reduction of a candidate for 3SAT to a
candidate for CHAIN can be done in polynomial time.  So we just need
to verify that there is a positive answer for CHAIN with $N=c$ if and
only if there is a positive answer for 3SAT.

If the boolean expression is satisfiable, then there is a variable
assignment that makes it true.  Consider the following interface
adapter chain.  In the variable handling subgraph, include edges that
correspond to the variable assignment.  In the clause handling
subgraph, there is guaranteed to be a path where all components
corresponding to clauses in the method availability vector at the
target end up being true, given the path in the variable handling
subgraph, so use this path in the chain.  Then $\|v^T\|$ will be
exactly $c$.

Conversely, suppose there is an adapter chain such that $\|v^T\|=c$.
Then assigning values to variables according to the path through the
variable handling subgraph results in a satisfying variable assignment
for the boolean expression.  This is because the clause handling
subgraph and the fact that $\|v^T\|=c$ together imply that all clauses
are true for the derived variable assignment.  And given an arbitrary
interface adapter chain and an optimal chain, it is easy to verify
whether the arbitrary adapter chain is not optimal, so CHAIN is
NP-complete.

\section{A greedy algorithm}
\label{sec:greedy-algorithm}

As shown in section~\ref{sec:complexity}, the problem of finding an
optimal interface adapter chain that would make available the most
methods in the target interface is an NP-complete problem.  Short of
developing a polynomial-time algorithm for an NP-complete problem,
practical systems will have to use a heuristic algorithm or an
exponential-time algorithm with reasonable performance in practice.

Algorithm~\ref{alg:greedy-algorithm} is a greedy algorithm that finds
an optimal interface adapter chain between a given source interface
and a target interface.  Given an interface adapter graph~$G$, it
works by looking at every possible acyclic adapter chain with an
arbitrary source that results in the target interface~$t$ in order of
increasing loss, taking advantage of
equation~(\ref{eq:numeric-monotonicity}), until we find a chain that
starts with the desired source interface~$s$.  In this context, loss
means the number of methods unavailable in the target interface given
a fully functional service with the source interface, which is
computed in algorithm~\ref{alg:loss}, so the algorithm is guaranteed
to find the optimal interface adapter chain.  In the worst case,
however, the algorithm takes exponential time since there can be an
exponential number of acyclic chains in an interface adapter graph.

\begin{algorithm}
  \caption{A greedy algorithm for interface adapter chaining.}
  \label{alg:greedy-algorithm}
  \begin{algorithmic}
  \Procedure{Greedy-Chain}{$G = (V, E)$, $s$, $t$}
    \State $\mathit{C} \gets \{ [] \}$
    \Comment{chains to extend}
    \State $\mathit{M} = \emptyset$
    \Comment{discarded chains}
    \State $D \gets \{[] \mapsto \mathbf{I}_{\dim(\mathbf{1}'_t)} \}$
    \Comment{method dependency matrixes}
    \While{$C \neq \emptyset$}
      \State $c \gets \mbox{element of $C$ minimizing $\textsc{Loss}(c,D)$}$
      \If{$c \neq [] \wedge \source{c[1]} = s$}
        \State \textbf{return} $c$
      \ElsIf{no acyclic chain not in $C \cup M$ extends $c$}
        \State $C \gets C - \{c\}$
        \State $M \gets M \cup \{c\}$
      \Else
        \If{$c = []$}
          \State $B \gets \{ [e] \,|\, e \in E, \target{e} = t\}$
        \Else
          \State $B \gets \{ e : c \,|\, e \in E, \target{e} = \source{c[1]} \}$
        \EndIf
        \State remove cyclic chains from $B$
        \State $C \gets C \cup B$
        \State $D \gets D \cup \{ e:c \mapsto D[c] \otimes \depend{e} \,|\, e : c \in B \}$
      \EndIf
    \EndWhile
  \EndProcedure
  \end{algorithmic}
\end{algorithm}

\begin{algorithm}
  \caption{Computing the lossiness of an interface adapter chain.}
  \label{alg:loss}
  \begin{algorithmic}
  \Function{Loss}{$c$, $D$}
    \State $s \gets \source{c[1]}$
    \State $t \gets \target{c[|c|]}$
    \State \textbf{return} $\dim(\mathbf{1}'_t) - \|D[c] \otimes \mathbf{1}'_s \|$
  \EndFunction
  \end{algorithmic}
\end{algorithm}

While algorithm~\ref{alg:greedy-algorithm} may take exponential time
in the worst case, results with a similar algorithm from
\cite{kim:percom2008} based on a small randomly generated interface
adapter graph suggest that the greedy algorithm has acceptable
performance in practice.

Algorithm~\ref{alg:greedy-algorithm} can easily be extended to support
the selection of an optimal source interface with weights associated
with methods expressing their importance as in
algorithm~\ref{alg:greedy-search}.  This can be done by checking that
the starting point of an interface adapter chain is included in a set
of possible source interfaces, instead of just comparing it to a
single source interface, and summing the weights for the available
methods in the target interface as in
algorithm~\ref{alg:weighted-adaptation} and using
equation~(\ref{eq:monotonicity}), instead of just counting the
methods.

\begin{algorithm}
  \caption{Greedy discovery for weighted interface adapter chaining.}
  \label{alg:greedy-search}
  \begin{algorithmic}
  \Procedure{Greedy-Chain}{$G = (V, E)$, $S$, $t$, $W$}
    \State $\mathit{C} \gets \{ [] \}$
    \Comment{chains to extend}
    \State $\mathit{M} = \emptyset$
    \Comment{discarded chains}
    \State $D \gets \{[] \mapsto \mathbf{I}_{\dim(\mathbf{1}'_t)} \}$
    \Comment{method dependency matrixes}
    \While{$C \neq \emptyset$}
      \State $c \gets \mbox{element of $C$ maximizing $\textsc{Weight}(c,D,W)$}$
        \If{$c \neq [] \wedge \source{c[1]} \in S$}
          \State \textbf{return} $(\source{c[1]}, c)$
      \ElsIf{no acyclic chain not in $C \cup M$ extends $c$}
        \State $C \gets C - \{c\}$
        \State $M \gets M \cup \{c\}$
      \Else
        \If{$c = []$}
          \State $B \gets \{ [e] \,|\, e \in E, \target{e} = t\}$
        \Else
          \State $B \gets \{ e : c \,|\, e \in E, \target{e} = \source{c[1]} \}$
        \EndIf
        \State remove cyclic chains from $B$
        \State $C \gets C \cup B$
        \State $D \gets D \cup \{ e : c \mapsto D[c] \otimes \depend{e} \,|\, e : c \in B \}$
      \EndIf
    \EndWhile
  \EndProcedure
  \end{algorithmic}
\end{algorithm}

\begin{algorithm}
  \caption{Computing the weight of an interface adapter chain.}
  \label{alg:weighted-adaptation}
  \begin{algorithmic}
  \Function{Weight}{$c$, $D$, $W = w_i$}
    \State $s \gets \source{c[1]}$
    \State $t \gets \target{c[|c|]}$
    \State $p_i \gets D[c] \otimes \mathbf{1}'_s$
    \State \textbf{return} $\sum_{p_i} w_i$
  \EndFunction
  \end{algorithmic}
\end{algorithm}

Unlike algorithm~\ref{alg:greedy-algorithm}, which would find an
interface adapter chain after a single service was presumably found by
a service discovery process, algorithm~\ref{alg:greedy-search} can be
used in the service discovery process itself to search for the best
service, not just in terms of what is required from the service, but
also considering how well the client could use the service.  And by
weighting the methods in the target interface, it can take into
account the importance of each method.  By having sufficiently large
weights for essential methods compared to those of non-essential
methods, algorithm~\ref{alg:greedy-search} can also guarantee that an
adapter chain which makes all essential methods available will always
be preferred over those which do not.

\section{Related work}
\label{sec:related-work}

The mathematics in this paper was motivated by the interface adapter
framework~\cite{kim:percom2008} used by the Active Surroundings
middleware for ubiquitous computing environments~\cite{lee:icat2004}.
In order to support a transparent computing experience despite a user
moving around locations where similar services may have different
interfaces, the framework uses interface adapters to adapt interfaces.
\cite{kim:percom2008} defines the problem informally and shows the
effectiveness of a greedy algorithm based on uniform cost
search~\cite{ucs}.

Other work have also used interface adapters to resolve service
interface mismatches.  Some attempt to aid developers create interface
adapters using template-based
approaches~\cite{benatallah:caise2005,nezhad} or mapping
specifications~\cite{zaha}.  Others reduce the number of required
interface adapters by chaining them together~\cite{ponnekanti}, while
others use a chain of interface adapters to provide backwards
compatibility as interfaces
evolve~\cite{hallberg:monadreader2005,kaminski:cascon2006}.  These
chaining approaches ignore that one chain may be worse than others in
terms of lossiness.

Analyzing the chaining of lossy interface adapters is in many ways
similar to depedency analysis in software
architecture~\cite{feldman:spe1979,loyall:csm1993,gao:noms2004,sangal:oopsla2005}.
These are designed to support maintenance of large software systems
and usually consider a lossy connection between software components to
be the exception and not the norm.  Techniques used in software
architecture such as code analysis~\cite{ryder:tse1979} or fault
injection~\cite{brown:im2001} could also be the basis for deriving the
method dependency matrixes for interface adapters.

\section{Conclusions}
\label{sec:conclusions}

By chaining a series of interface adapters, it is possible for a
single-interface client use a much wider variety of services with
heterogeneous interfaces without requiring an explosive number of
interface adapters.  However, as an interface adapter may not be able
to convert one interface to another perfectly, we have developed a
mathematical framework which can be used to analyze the lossiness
incurred during chained interface adaptation.

The mathematical framework defines the method dependency matrix, the
method availability vector, and the composition operation for
describing the properties of composed adapters, which was also proved
to be associative.  The framework could be used to analyze the
lossiness in interface adapter chains and develop algorithms for
finding such chains.

However, finding an optimal interface adapter chain is an NP-complete
problem, which can be proven by reducing 3SAT to CHAIN.  A greedy
algorithm for finding an optimal interface adapter chain requiring
exponential time in the worst case was suggested.

This paper has only considered the all-or-nothing case where a method
in a target interface can be completely implemented using methods in
the source interface.  However, in certain cases the method could only
be implemented partially.  One possible extension to the mathematical
framework is to consider partial adaptation of such methods.
Extending it so that it can analyze the lossiness when services are
composed is another possibility.  Heuristic algorithms with provable
approximation bounds is another topic that would be worth looking into
in the future.

\bibliographystyle{plain}
\bibliography{strings,local,articles,proceedings}

\end{document}